\documentclass[aps,prx,groupaddress,showpacs,twocolumn]{revtex4}
\usepackage{graphicx,bm,amsmath,amsfonts,bbold, amssymb,color,microtype}

\newcommand{\be}{\begin{equation}}
\newcommand{\ee}{\end{equation}}
\newcommand{\bea}{\begin{eqnarray}}
\newcommand{\eea}{\end{eqnarray}}
\newcommand{\im}{\mbox{Im}}

\newcommand{\cs}[1]{\cos(#1\phi)}
\newcommand{\s}[1]{\sin(#1\phi)}

\newcommand{\wf}{\varphi}
\newcommand{\twf}{\tilde{\varphi}}
\newcommand{\tk}{\tilde{k}}
\newcommand{\ttau}{\tilde{\tau}}

\newcommand{\e}{\epsilon}
\newcommand{\ttauEP}{\ttau_{\!\scriptscriptstyle \text{TH}}}
\newcommand{\tauEP}{\tau_{\!\scriptscriptstyle \text{TH}}}

\begin{document}

\title{Parity-Time Symmetry Breaking beyond One Dimension: the Role of Degeneracy}

\author{Li Ge}
\email{li.ge@csi.cuny.edu}
\affiliation{\textls[-18]{Department of Engineering Science and Physics, College of Staten Island, CUNY, Staten Island, NY 10314, USA}}
\affiliation{The Graduate Center, CUNY, New York, NY 10016, USA}
\author{A. Douglas Stone}
\affiliation{Department of Applied Physics, Yale University, New Haven, CT 06520-8482, USA}

\begin{abstract}
We consider the role of degeneracy in Parity-Time ($\cal{PT}$) symmetry breaking for non-hermitian wave equations beyond one dimension. We show that if the spectrum is degenerate in the absence of $\cal T$-breaking, and $\cal T$ is broken in a generic manner (without preserving other discrete symmetries), then the standard $\cal{PT}$-symmetry breaking transition does not occur, meaning that the spectrum is complex even for infinitesimal strength of gain and loss.
However the realness of the entire spectrum can be preserved over a finite interval if additional discrete symmetries $\chi$ are imposed when $\cal T$ is broken, if $\chi$ decouple all degenerate modes. When this is true only for a subset of the degenerate spectrum, there can be a partial $\cal PT$ transition in which this subset remains real over a finite interval of $\cal T$-breaking. If the spectrum has odd-degeneracy, a fraction of the degenerate spectrum can remain in the symmetric phase even without imposing additional discrete symmetries, and they are analogous to dark states in atomic physics. These results are illustrated by the example of different $\cal T$-breaking perturbations of a uniform dielectric disk and sphere, and a group theoretical analysis is given in the disk case.  Finally, we show that multimode coupling is capable of restoring the $\cal PT$-symmetric phase at finite $\cal T$-breaking. We also analyze these questions when the parity operator is replaced by another spatial symmetry operator and find that the behavior can be qualitatively different.
\end{abstract}

\maketitle

\section{introduction}


Parity-Time ($\cal{PT}$) symmetric systems have attracted considerable interest in the past few years. These are non-hermitian systems which are invariant under the combined action of a parity and time-reversal operation.  In the case of closed Hamiltonian systems the transition is from a regime of real energy eigenvalues to complex conjugate pairs of eigenvalues as the degree of non-hermiticity is increased \cite{Bender1,Bender2,Bender3}.  For the case of open, scattering systems, the transition is seen in the eigenvalues of the scattering matrix, which can remain on the unit circle despite the non-hermiticity up to some threshold and then depart from it in pairs with inverse moduli \cite{CPALaser,conservation,Robin}.  In both cases the transition occurs when two eigenvalues coincide at an exceptional point (EP) which corresponds not to a degeneracy of the relevant operator but to a point at which it becomes defective (two eigenvectors coalesce), and hence is non-diagonalizable \cite{EP1,EP2,EPMVB,EP3,EP4,EP5,EP6,EP7}.  A major application of the theory of $\cal PT$-symmetry breaking is to the wave equation of electromagnetism where the possibility of adding gain and loss in a $\cal PT$-symmetric manner allows observation of many intriguing phenomena \cite{El-Ganainy_OL06, Moiseyev,Kottos,Musslimani_prl08,Makris_prl08,Longhi,CPALaser,conservation,Robin,RC,Microwave,Regensburger,Kottos2}.

Essentially all of the work on $\cal PT$-symmetry breaking has focused on one-dimensional (1D) or quasi-1D (coupled waveguide) systems. These systems can never have a high enough symmetry group to generate generic degeneracies. In the current work we will focus on two-dimensional (2D)  and three-dimensional (3D) $\cal PT$-symmetric wave equations, which can have the new feature of continuous symmetries and generic degeneracies in the absence of the $\cal T$-breaking non-hermitian perturbation. It will be shown that for such systems the $\cal PT$-transition is absent if $\cal T$ is generically broken, meaning that they do not have a real spectrum even when the $\cal T$-breaking is \textit{infinitesimal}.
However, if $\cal T$ is not generically broken, i.e. if some further discrete spatial symmetries are preserved, then it is possible that either the entire spectrum remains real over a finite interval (standard $\cal PT$ behavior) or a finite subset of the degenerate spectrum does. These scenarios are analyzed using a coupled-mode theory and generalized point groups. Our analysis also shows that other composite symmetries which can occur in higher dimension, such as $\cal RT$ where $\cal R$ represents rotation by $\pi$, can behave differently from $\cal PT$ and can exhibit a fully real spectrum when the corresponding $\cal PT$ system does not.
Finally, we show that it is also possible for multimode coupling to restore the $\cal PT$-symmetric phase, at finite $\cal T$-breaking, if it is appropriately tuned.

\section{role of degeneracy: a qualitative argument}

We now give a qualitative argument for the absence of $\cal PT$-transition due to degeneracy, which will be quantified later using a coupled-mode theory.
We focus on the case of closed, hamiltonian systems, $H$; analogous conclusions should hold for scattering systems with unimodularity of the eigenvalues replacing realness \cite{smatrix_deg}. Since the hamiltonian has $\cal PT$-symmetry, it is easily shown that the eigenfunctions of $H$ have closure under the $\cal PT$ operation: in the symmetric ``phase" each eigenfunction is mapped to itself; in the broken symmetry phase it is mapped to another in a finite set.  Consider the non-degenerate case. When there is no $\cal T$-breaking, then the system is hermitian and the eigenvalues, $\{k_j\}$, are real.  We will use a real parameter $\tau$ to denote the strength of the non-hermitian, $\cal T$-breaking term in the hamiltonian.
For a particular eigenvalue $k_j$, it must move continuously as $\tau$ is increased from zero, because there are no singularities in the problem which would allow a jump (the derivative of the eigenvalue motion can be singular, e.g. at an exceptional point, but finite jumps never occur).
Given this constraint, is it possible for $k_j$ to make a small step off the real axis?  The answer is no, its movement must be confined to the real axis in the vicinity of $\tau = 0$.  The reason is
that the $\cal PT$-symmetry of $H$ implies that under $\cal PT$ this eigenvalue is always mapped to its complex conjugate, which is trivially satisfied if $k_j$ remains real.  But if $k_j$ moved off the real axis, then another
eigenvalue would have to move off as well so as to form a complex conjugate pair with the same real part and opposite imaginary parts for
this infinitesimal value of $\tau$. But by the generic non-degeneracy assumption,
all other eigenvalues are a finite distance away on the real axis and cannot possibly move enough to satisfy the symmetry as $\tau \to 0$.  This explains why the $\cal PT$-symmetry breaking point requires a finite $\tau$
for non-degenerate hamiltonians (and hence all hamiltonians in finite 1D systems).  In both the hamiltonian case and the scattering case, pairs of eigenvalues must move a finite distance on the real axis or unit circle before they meet (at an EP) in order to undergo the transition while satisfying the exact $\cal PT$-symmetry at all values of $\tau$.

Now consider a higher dimensional case with generic even degeneracy (the odd degeneracy, which can occur in 3D, will be discussed below).  In this case the previous argument does not hold. As the $\cal T$-breaking is turned on, it is possible for the paired eigenvalues to have a first-order splitting in $\tau$ while maintaining the required $\cal PT$-symmetry, and indeed this becomes the generic case, and the $\cal PT$-symmetric phase no longer occurs. In this case the degeneracy can be seen as the precursor of the $\cal PT$-broken phase but no eigenfunction coalesce occurs as it would at an EP. We show rigorously some examples of this behavior below using a coupled-mode theory, and analyze the role of additional discrete symmetries and multimode coupling as means to protect or restore the $\cal PT$-symmetric phase.

While the above argument is qualitative at this level, it is based on the rigorous requirement that if $k_j$ is an eigenvalue of a ${\cal PT}$ symmetric $H$, then so must be $k_j^*$, and the fact that the roots of a characteristic polynomial must be a continuous functional of the complex potential in the wave equation. Hence it should be straightforward to build a rigorous proof of the following statement: the existence of degeneracy at $\tau=0$ is a {\it necessary condition} for thresholdless $\cal PT$ breaking (transition for infinitesimal $\tau$). A second statement that also follows is: degeneracy is also a {\it sufficient condition} for a thresholdless transition if the $\cal T$-breaking perturbation couples the degenerate modes of the $\tau = 0$ system.  We will not attempt rigorous mathematical proofs of these statements but will provide arguments supporting them below.

\section{Quantitative study of 2D disk: even degeneracy}

First we consider 2D circular systems of radius $R$, in which the eigenmodes $\twf(\vec{r})$ are determined by the Helmholtz equation
\be
-\nabla^2 \twf(\vec{r}) = [\e_c(\vec{r})+i\tau g(\vec{r})]\frac{\tilde{\omega}^2}{c^2}\twf(\vec{r}). \label{eq:H}
\ee
We refer to $\tk=\tilde\omega/c$ as the eigenfrequency of the system and employ the Dirichlet boundary condition.
The cavity dielectric function $\e_c(\vec{r})$, gain and loss strength $\tau$, and their spatial profile $g(\vec{r})$ are real quantities. We adopt the convention that $\tau$ is non-negative, with which $g(\vec{r})<0$ $(>0)$ represents gain (loss). We take a uniform $\e_c(\vec{r})=n^2$ which results in a separable system at $\tau=0$; each eigenmode of Eq.~(\ref{eq:H}) has a well defined angular momentum $m$, and a second quantum number $\eta$ indicates the number of intensity peaks in the radial direction. Henceforth we denote ${\cal P}_\phi$ as the parity operation about the $\phi,\phi+\pi$ axis, and we choose ${\cal P}\equiv{\cal P}_0$ as the parity operation in the $\cal PT$-symmetry, i.e. $g(r,\phi)=-g(r,-\phi)$. Except for the $m=0$ modes, the eigenmode spectrum at $\tau=0$ consists of degenerate pairs due to the rotational symmetry. The eigenmodes can be conveniently expressed as $\wf_{m,\eta}^{(p)}(\vec{r})\propto J_m(n k_{m,\eta} r)\cs{m}, J_m(n k_{m,\eta} r)\s{m}$, and we note that the corresponding quantities with and without the tilde are defined in general and at $\tau=0$, respectively. $\{k_{m,\eta}\}$ are real and the superscript $p$ indicates even ($e$) or odd ($o$) parity about the $\cal PT$-axis.
The eigenmodes at $\tau=0$ form an orthogonal and complete basis, satisfying $\int d\vec{r} \wf_j(\vec{r})\wf_{j'}(\vec{r})=\delta_{jj'}$, in which we have used a single index $j$ to represent $\{m,\eta,p\}$.

At $\tau\neq0$ the system becomes non-hermitian, and its eigenmodes $\twf(\vec{r})$ can be expanded in the hermitian basis $\{\wf_j(\vec{r})\}$, i.e. $\twf(\vec{r}) = \sum_j a_j \wf_j(\vec{r})$. For the superposition of two $\wf_j$'s  to be in the $\cal PT$-symmetric phase, i.e. $\twf(\vec{r}) \propto \twf^*(\mathcal{P}\vec{r})$, it requires either (i) the $\wf_j$'s have the same parity and $a_j$'s are in phase; or (ii) they have opposite parity and $a_j$'s are $\pi$-out-of-phase.
Condition (i) implies that $\twf(\vec{r})$ is still real with a proper normalization, while condition (ii) requires a complex $\twf(\vec{r})$. These conditions can be easily generalized if $\twf(\vec{r})$ contains more than two $\wf_j$'s.

\begin{figure}[t]
\begin{center}
\includegraphics[width=\linewidth]{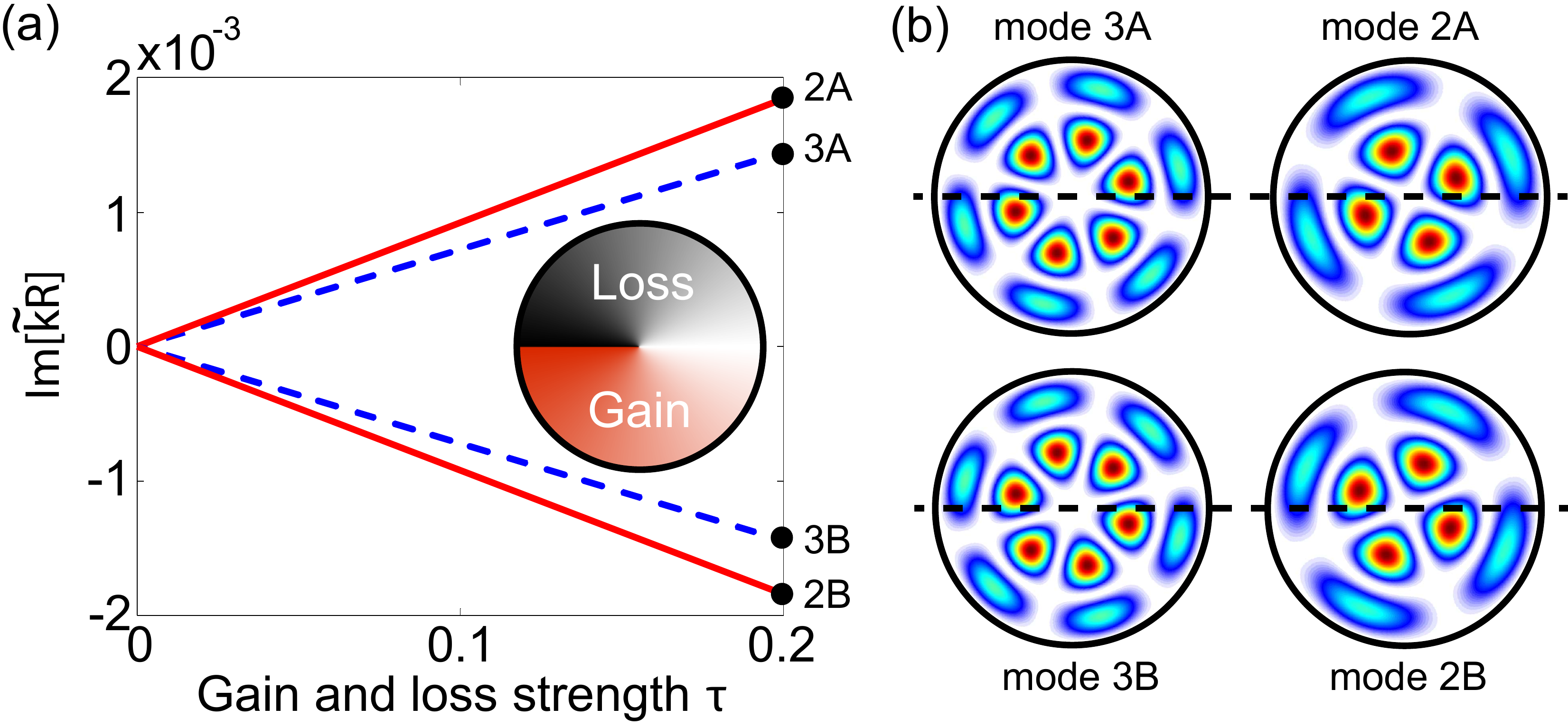}
\caption{(Color online) Absence of $\cal PT$ transitions in a $\cal{PT}$-symmetric disk with no additional discrete symmetry. The gain and loss gradient is given by $g(r,\phi)=\phi/\pi (r>R/5;-\pi<\phi<\pi), 0(r\leq R/5)$ and shown schematically as the inset in (a).
(a) Evolution of the imaginary part of two pairs of initially degenerate eigenfrequencies as a function of $\tau$. Solid and dashed lines represent modes with a dominant angular momentum $m=2,3$, respectively. These numerical results cannot be distinguished from the approximation (\ref{eq:imkR}). (b) Intensity profiles of the four marked modes in (a) at $\tau=0.2$. Dashed lines show the designated $\cal P$-axis. The refractive index $n=3.3$ is used in all examples.}
\label{fig:linear}
\end{center}
\end{figure}

\subsection{Absence of $\cal PT$ transitions}

We analyze the spectrum of disks with generically broken $\cal T$ symmetry, and consistent with our
argument in Section II we find that the previously degenerate modes (at $\tau=0$) are now in the $\cal PT$-broken phase at an {\it arbitrarily} small $\tau$.  Generically broken means that the only residual symmetry
of the circle maintained in the non-hermitian system is the designated $\cal PT$ symmetry. One example of such a case is shown in Fig.~\ref{fig:linear}, where the gain and loss profile increases linearly from $\phi=0$ to $\mp\pi$, retaining $\cal PT$ but breaking all other combinations of a discrete symmetry and $\cal T$. The two pairs of eigenfrequencies shown have a dominant angular momentum $m=2,3$, respectively. For each pair, their frequencies are complex conjugates, as is demanded by $\cal PT$ symmetry, and their intensity profiles are mirror images of each other about the $\cal PT$-axis;
the one that overlaps strongly with the gain region in each pair has a positive $\im[\tk]$, and it increases linearly with $\tau$.

To understand quantitatively the absence of the standard $\cal PT$-phase transition at a finite $\tau$, we develop a coupled-mode theory using the hermitian modes at $\tau=0$ as the basis:
\be
\left[\mathbb{1} + i\ttau\bm{G}\right]\bm{a} = \tk^{-2} \bm{E} \bm{a}.\label{eq:modeCoupling}
\ee
In this matrix equation $\ttau\equiv\tau/n^2$, $G_{jj'}\equiv\int d\vec{r} g(\vec{r})\wf_j(\vec{r})\wf_{j'}(\vec{r})$ is the coupling between any two hermitian modes, $\bm{E}_{jj'} =k_j^2 \delta_{j,j'}$ is a diagonal matrix
with the squared eigenvalues on the diagonal arranged in degenerate pairs; $\bm{a}$ is the column vector of the expansion coefficients in the $\{\wf_j(\vec{r})\}$ basis. Note that the coupled-mode theory (\ref{eq:modeCoupling}) is an exact restatement of  Eq.~(\ref{eq:H}) if all modes of the $\tau=0$ system are
taken into account. Our approach here is different from the previous types of coupled-mode theory applied in $\cal PT$-symmetric systems (see, for example, Ref.~\cite{El-Ganainy_OL06}), which consider the coupling between two eigenmodes of the {\it subsystems}, e.g. a waveguide with gain and a neighboring one with loss. We on the other hand, focus on the couplings between modes of the {\it entire system} at $\tau =0$ \cite{Moiseyev}. In the first part of the paper we will focus on the case in which (\ref{eq:modeCoupling}) is truncated to degenerate sets of basis functions, because only for these
is the effect of the the ${\cal T}$-breaking first order in $\tau$.  Later we will use (\ref{eq:modeCoupling})
to discuss multimode couplings beyond the degenerate sets.

For the degenerate pairs in the disk Eq.~(\ref{eq:modeCoupling}) is simply
\be
\begin{pmatrix}
1  & i\ttau G_{eo}\\
i\ttau G_{eo} & 1
\end{pmatrix}
\begin{pmatrix}
a_e\\
a_o
\end{pmatrix}
=
\frac{k_j^2}{\tk^2}
\begin{pmatrix}
a_e\\
a_o
\end{pmatrix},\label{eq:modeCoupling2x2}
\ee
and $G_{ee},G_{oo}$ on the diagonal vanishes due to the $\cal PT$-symmetry. We immediately find that
\be
\im[\tk] \approx \pm\frac{k_jG_{eo}}{2}\ttau, \label{eq:imkR}
\ee
which agrees nicely with the linear $\tau$-dependence of $\im[\tk]$ shown in Fig.~\ref{fig:linear}.
As expected from first-order perturbation theory, the rate at which the complex splitting of the eigenfrequencies increases depends on the gain and loss profile $g(\vec{r})$, while the mixing ratio between each pair of degenerate modes does not. This can be directly seen from the $2\times2$ matrix on the left hand side of Eq.~(\ref{eq:modeCoupling2x2}), whose diagonal and off-diagonal elements are the same, leading to the usual symmetric and anti-symmetric superpositions
$\twf_{j,\pm}=(\wf^{(e)}_j\pm\wf^{(o)}_j)/\sqrt{2}$, which can be easily checked to be $\cal PT$-symmetric partners, satisfying $({\cal PT})\twf_{j,\pm}=\twf_{j,\mp}$. Alternatively, we could have used the degenerate pair $\wf_{j,\pm}\equiv\twf_{j,\pm}$ in the hermitian basis instead of $\wf^{(e)}_j$ and $\wf^{(o)}_j$, which has no off-diagonal elements in the perturbation, and it would have led to the same expression (\ref{eq:imkR}), but for the subsequent analysis it is useful to work in the parity basis.

To illustrate the crossover to the non-degenerate case where there is a $\cal PT$-phase transition, we analyze Eq.~(\ref{eq:modeCoupling}) for two non-degenerate eigenvalues $k_j, k_{j'}$, assuming that they are relatively close so that couplings to other eigenstates can be ignored. The two resulting eigenvalues due to the coupling of $\wf_j$ and $\wf_{j'}$ are given by
\be
\tk^{2} = \frac{2k_j^{2}k_{j'}^{2}}{(k_j^{2}+k_{j'}^{2})\pm\sqrt{(k_j^{2}-k_{j'}^{2})^2-4\ttau^2{G_{jj'}}^2k_j^{2}k_{j'}^{2}}}, \label{eq:modeCoupling2x2b}
\ee
which are necessarily real for a small $\ttau$ if $k_j\neq k_{j'}$. The transition occurs when $\ttau$ becomes larger than
\be
\ttauEP = \frac{|k_j^{2}-k_{j'}^{2}|}{2|G_{jj'}|k_jk_{j'}}\approx\frac{|k_j-k_{j'}|}{|G_{jj'}|k_j}\label{eq:TH}.
\ee
It is clear that $\ttauEP$ approaches zero as $k_j\rightarrow k_{j'}$, and we recover Eq.~(\ref{eq:imkR}) from Eq.~(\ref{eq:modeCoupling2x2b}) in this limit when $\ttau|G_{jj'}|\ll1$.
In the absence of degeneracy or quasi-degeneracy, multimode coupling will typically be important, and the $\cal PT$ transition strength cannot be quantitatively determined by the above expression (\ref{eq:modeCoupling2x2b}).

\subsection{Protected $\cal PT$-symmetric phase due to additional discrete symmetries}

As noted above, the arguments for a thresholdless transition hold for generic $\cal T$-breaking perturbations, and they can break down for cases in which imposition of gain and loss preserves additional discrete symmetries $\chi$.  In such cases the condition for thresholdless $\cal PT$ breaking for a pair of degenerate modes is that their coupling $G_{eo}$ must be finite. In the most extreme case, $\chi$ may decouple {\it all} degenerate pairs, i.e. all the corresponding $G_{eo}$ in Eq.~(\ref{eq:modeCoupling}) vanish, and a finite $\cal PT$-transition threshold $\ttauEP$ for the entire spectrum is restored, which is due solely to the coupling between non-degenerate modes.

\begin{figure}[t]
\begin{center}
\includegraphics[width=\linewidth]{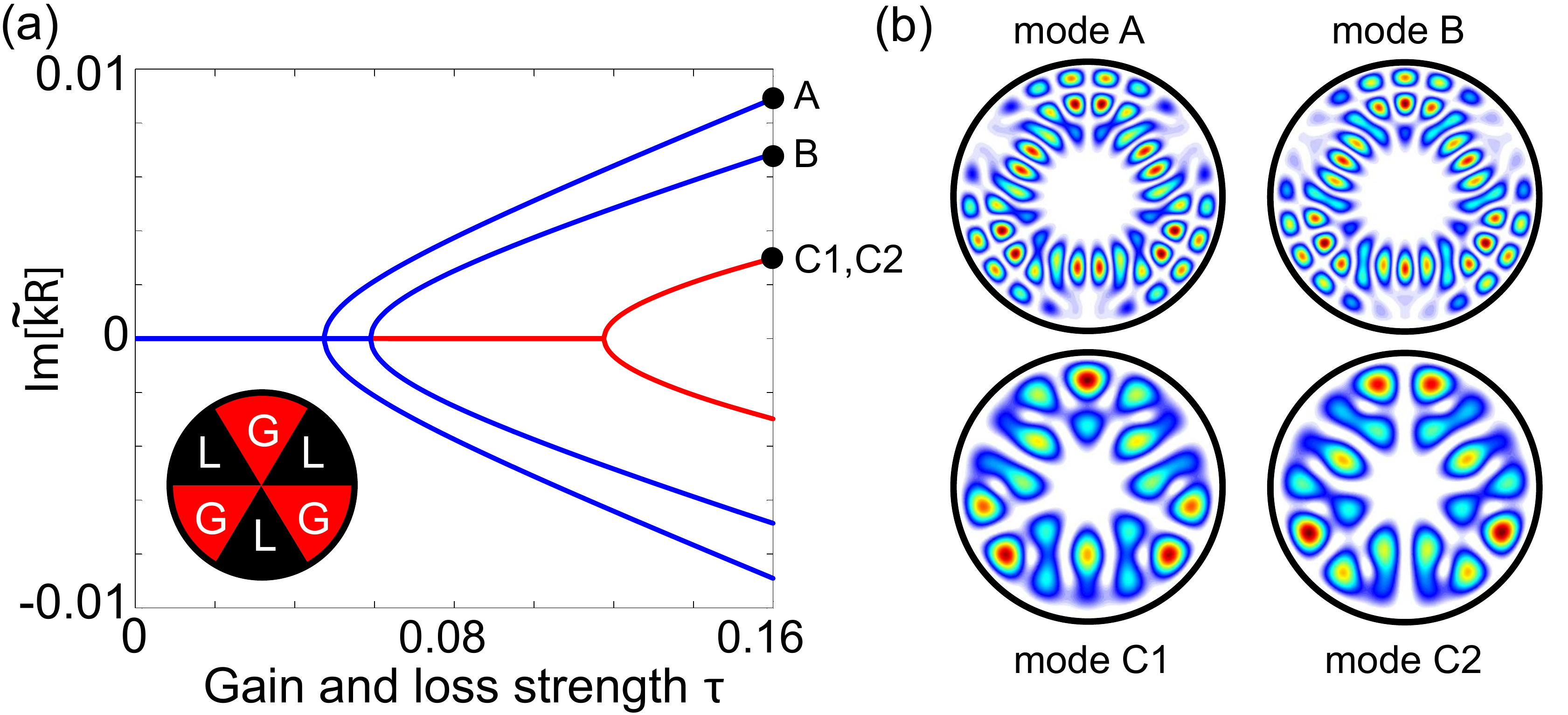}
\caption{(Color online) Protected $\cal PT$-transitions and an entirely real spectrum at small $\tau$ with additional discrete symmetries. (a) Three pairs of broken-symmetry modes are shown, and complex eigenvalues only appear above the lowest $\tauEP\simeq0.047$.
Inset: the symmetries of the system is described by the generalized dihedral group ${\cal DT}_{2v}$ with $v=6$ (see Section~\ref{sec:group}). It includes, for example, ${\cal P}_{\frac{\pi}{2}}$ and ${\cal R}_{\frac{2\pi}{3}}$ in addition to the $\cal PT$-symmetry. Here ${\cal P}_\phi$ denotes the parity operation about the $\phi,\phi+\pi$ axis, and ${\cal R}_\phi$ denotes clockwise rotation about the origin by $\phi$.
(b) Intensity profiles of the four broken-symmetry modes marked in (a) at $\tau=0.16$. The broken-symmetry modes can be non-degenerate (e.g. mode A and B), in which case they are simultaneous eigenfunctions of ${\cal P}_{\frac{\pi}{2}}$ and ${\cal R}_{\frac{2\pi}{3}}$; they can also be degenerate (e.g. mode C1 and C2), in which case they can either be eigenfunctions of ${\cal P}_{\frac{\pi}{2}}$ or ${\cal R}_{\frac{2\pi}{3}}$ but not both. Mode C1 and C2 plotted are eigenfunctions of ${\cal P}_{\frac{\pi}{2}}$, and they can be linearly combined to be eigenfunctions of ${\cal R}_{\frac{2\pi}{3}}$ \cite{2deg}. }
\label{fig:radiationHazard}
\end{center}
\end{figure}

While the results we have obtained above follow straightforwardly from the perturbative coupled-mode approach, they have not been anticipated by earlier work; we believe this is due to the relatively simple $\cal T$-breaking perturbations typically imposed
in higher dimensions, which {\it do} preserve discrete symmetries. Such an example occurred in our own work \cite{CPALaser}, in which the scattering $\cal PT$ transitions were considered in a disk consisting of two uniform gain and loss halves. The protection of the symmetric phase for the entire spectrum at small $\tau$ can be attributed to $\chi\equiv{\cal P}_{\frac{\pi}{2}}$ in this case. This is because the integrand of $G_{eo}$ for each degenerate pair is an odd function with respect to ${\cal P}_{\frac{\pi}{2}}$, and thus all these $G_{eo}$ vanish as a result. This observation is confirmed for more complicated $g(\vec{r})$ satisfying the ${\cal P}_{\frac{\pi}{2}}$ symmetry, including the ``radiation hazard'' configuration shown in Fig.~\ref{fig:radiationHazard}.

\begin{figure}[t]
\begin{center}
\includegraphics[width=\linewidth]{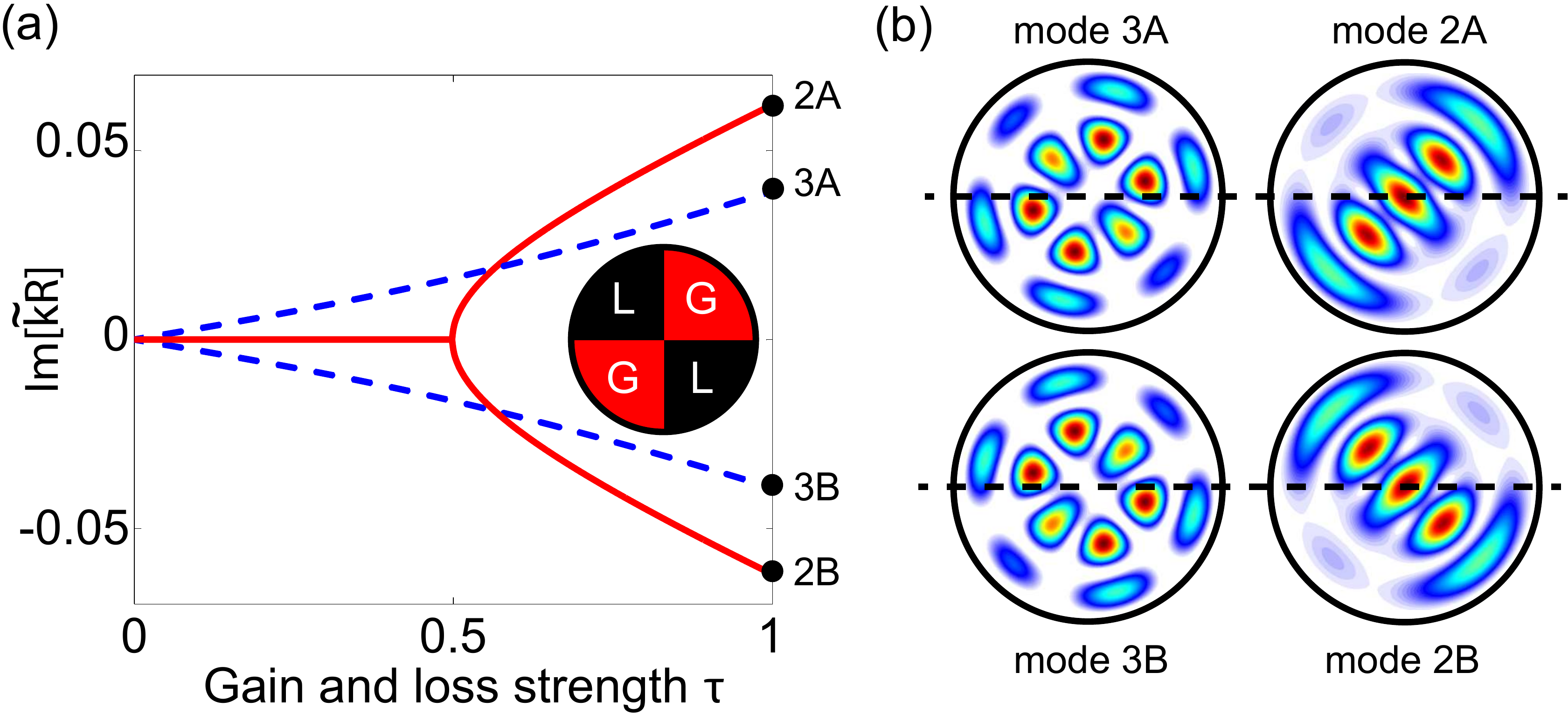}
\caption{(Color online) Partial transitions with additional discrete symmetries. (a) Same as Fig.~\ref{fig:linear} for the $m=2,3$ modes but with uniform gain or loss in each of the four quadrants, which is described by the generalized dihedral group ${\cal DT}_{2v}$ with $v=4$ (see Section~\ref{sec:group}). Dashed lines show the absence of the standard $\cal PT$-phase transition for the $m=3$ modes.
Solid lines show the standard $\cal PT$-phase transition for the odd-parity $m=2$ mode coupled to a nearby $m=0$ mode. Their threshold $\tauEP=0.498$ is well approximated by Eq.~(\ref{eq:TH}) which gives $\tauEP=0.483$. The pairing even-parity $m=2$ mode stays in the $\cal PT$-symmetry phase in the range shown.
(b) Intensity profiles of the four modes at $\tau=1$. The $m=3$ modes are slightly perturbed from those in Fig.~\ref{fig:linear} due to their weak coupling to a pair of $m=1$ modes nearby as $\tau$ becomes larger.}
\label{fig:PTwheel}
\end{center}
\end{figure}

The situation discussed above, i.e. $\chi$ decouples all degenerate modes, is a special scenario. More generally, $\chi$ only decouples a subset of the degenerate modes, and a partial $\cal PT$-breaking transition occurs as a result. One such example, the ``$\cal PT$-wheel'' configuration, is shown in Fig.~\ref{fig:PTwheel}. It has a uniform gain and loss profile in each of the four quadrants, invariant with the ${\cal P}_{\pm\frac{\pi}{4}}$ operations. For two degenerate modes with an even $m$, the integrand of the corresponding $G_{eo}$ is an odd function with respect to ${\cal P}_{\pm\frac{\pi}{4}}$. Therefore, $G_{eo}$ vanishes and these modes are protected from symmetry breaking at infinitesimal $\tau$; $\wf_j^{(e)},\wf_j^{(o)}$ need to couple to other more distant modes to break the $\cal PT$-symmetry, which is the standard scenario described by Eq.~(\ref{eq:modeCoupling2x2b}) with a finite threshold $\tauEP$ [solid lines in Fig.~\ref{fig:PTwheel}(a)]. On the other hand, the integrand of $G_{eo}$ is an even function with respect to ${\cal P}_{\pm\frac{\pi}{4}}$ for two degenerate modes with an odd $m$. Therefore, $G_{eo}$ {\it does not vanish} in this case and any pair of odd-$m$ degenerate modes are in the $\cal PT$-broken phase with infinitesimal $\tau$ [dashed lines in Fig.~\ref{fig:PTwheel}(a)].
We note that a similar partial transition in coupled waveguides was mentioned by \u{C}tyrok\'y in Ref.~\cite{Ctyroky}.

\subsection{General symmetry analysis}
\label{sec:group}
The arguments and examples above support our general conjectures that degeneracy is a necessary but not sufficient condition for a thresholdless $\cal PT$ transition.  As shown in the example of Fig. 1, when the
$\cal T$ breaking perturbation preserves no discrete symmetries (i.e. is generic) then degeneracy is sufficient, but otherwise the situation must be analyzed separately depending on the nature of the remaining symmetries $\chi$ in the presence of $\cal T$ breaking.  We will now do this
using the generalized ``point group" ${\cal S}\equiv\{{\cal PT},\chi\}$, where $\chi$ also includes the identity operator $\mathbb 1$.


The examples examined above and in Ref.~\cite{Ctyroky} have $v$ angular blocks of equal area, and the corresponding $\cal S$ is a generalization of the dihedral group, $D_{2v}$, describing the symmetries of regular polygons of $v$ sides. $D_{2v}$ contains $2v$ elements, including $\mathbb 1$, $v-1$ rotations, and $v$ reflections.
We denote the relevant generalization of this group to our system as ${\cal DT}_{2v}$.
We see that due to the $\cal PT$-symmetry, $v$ needs to be an even number.
The difference between ${\cal DT}_{2v}$ and $D_{2v}$ is due to the effect of $\cal T$ breaking while $\cal PT$ is preserved. One finds that $v$ elements in $D_{2v}$ are no longer symmetry operators, but become so again when multiplied by $\cal T$.
In the simplest example, a disk with half uniform gain and loss considered in Ref.~\cite{CPALaser}, ${\cal S}={\cal DT}_4\equiv\{\mathbb{1},{\cal PT},{\cal P}_\frac{\pi}{2},{\cal R}_{\pi}{\cal T}\}$, where ${\cal R}_\phi$ denotes clockwise rotation about the origin by $\phi$. In comparison, $D_{4}$ contains $\mathbb{1},{\cal P},{\cal P}_\frac{\pi}{2}$, and ${\cal R}_{\pi}$. Similarly, for the ``$\cal PT$-wheel" configuration shown in Fig.~\ref{fig:PTwheel}, ${\cal S}={\cal DT}_8\equiv\{\mathbb 1, {\cal PT}, {\cal P}_{\pm\frac{\pi}{4}}, {\cal P}_\frac{\pi}{2}{\cal T}, {\cal R}_\frac{\pi}{2}{\cal T}, {\cal R}_{\pi}, {\cal R}_\frac{3\pi}{2}{\cal T}\}$, and removing the 4 $\cal T$-operators gives the original $D_8$ group. Although in these examples each angular block only contains gain or loss, this does not need to be the case. For example, the group property is not affected if we exchange gain and loss on a ring centered at the origin.


We want to find the conditions for vanishing coupling $G_{eo}$ between pairs of degenerate eigenfunctions, which will then lead to a protected $\cal PT$ transition for that set or subset of the
states.  Assume $\wf_j^{(e)},\wf_j^{(o)}$ are a pair of degenerate eigenfunctions of $H$ and they are also
eigenfunctions of at least one element of $\cal S$ (denoted by ${\cal S}_v$). It is easy to see that their eigenvalues for ${\cal S}_v$ can only be either 1 or -1.  If ${\cal S}_v$ contains $\cal T$ and this pair are degenerate in terms of ${\cal S}_v$, or
if ${\cal S}_v$ does not contain $\cal T$ and they are non-degenerate in terms of ${\cal S}_v$, it follows that the integrand of $G_{eo}$ is then an odd function with respect to ${\cal S}_v$, and $G_{eo}$ vanishes.

${\cal P}_\frac{\pi}{2}$ is such an operator ${\cal S}_v$ for all pairs of $\wf_j^{(e)},\wf_j^{(o)}$ as indicated before, and so is ${\cal R}_{\pi}{\cal T}$.
The presence of either of these operators is sufficient to protect the entire spectrum from the effects of infinitesimal $\cal T$ breaking and restore the normal $\cal PT$ transition.
For the generalized dihedral group ${\cal DT}_{2v}$ this criterion is satisfied when $v$ is the sum of 2 and a multiple of 4, and both operators ${\cal P}_\frac{\pi}{2}$, ${\cal R}_{\pi}{\cal T}$ appear (due to the closure property of a group given ${\cal R}_{\pi}{\cal T} = {\cal P}_\frac{\pi}{2}{\cal PT}$).  Examples of this type were considered in Ref.~\cite{CPALaser} (${\cal DT}_{4}$) and here in Fig.~\ref{fig:radiationHazard} (${\cal DT}_{12}$). For the rest, i.e. ${\cal DT}_{2v}$ with $v$ that is a multiple of 4 (such as ${\cal DT}_{8}$ in Fig.~\ref{fig:PTwheel}), they do not include either ${\cal P}_\frac{\pi}{2}$ or ${\cal R}_{\pi}{\cal T}$, and there are always some $G_{eo}$ that do not vanish, and an entirely real spectrum cannot be maintained at infinitesimal $\cal T$-breaking.

It is interesting to note that these two different scenarios with the generalized dihedral group can be differentiated by the presence or absence of  $\cal RT$-symmetry (where $\cal R\equiv R_\pi$ is rotation by $\pi$ and equivalent to inversion in 2D systems). It has been speculated that parity in $\cal PT$-symmetric structures can be replaced by {\it any} linear symmetry operator, such as rotation or inversion, and the same spectral phase transition behavior will be observed as for $\cal PT$. Here, when we take into account the role of degeneracy, we find that $\cal PT$ systems can have qualitatively different behavior from $\cal RT$.  Whereas $\cal RT$ always give a finite $\tau$ transition for the full spectrum, $\cal PT$ can give all three behaviors: finite $\tau$, infinitesimal $\tau$ and a mixture of both.

\begin{figure}[t]
\begin{center}
\includegraphics[width=\linewidth]{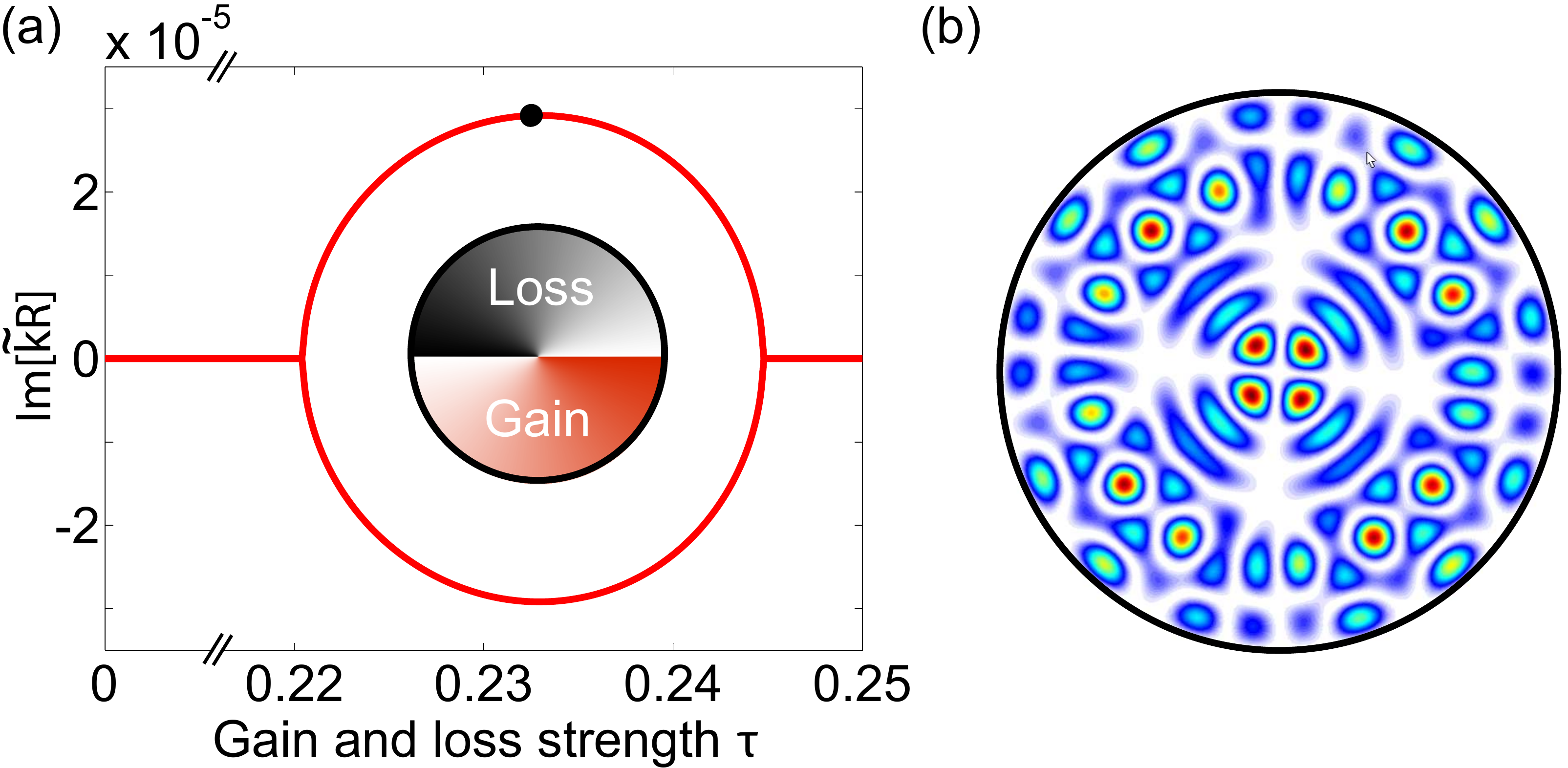}
\caption{(Color online) A finite transition threshold always appears in a $\cal{RT}$-symmetric disk, with or without additional symmetries. The one shown in the inset of (a) belongs to the former, and the gain and loss gradient is given by $g(r,\phi)=\phi/\pi (r>R/5;0<\phi<\pi), -(\phi+\pi)/\pi (r>R/5;-\pi<\phi<0), 0 (r\leq R/5)$.
In (a) only the two eigenfrequencies with the lowest $\tauEP$ are shown. (b) shows the intensity profile of the marked mode at $\tau=0.233$. The intensity profile of its $\cal RT$-symmetric partner is obtained by rotating (b) by $\pi$.}
\label{fig:linear_RT}
\end{center}
\end{figure}

This statement about $\cal RT$ generalizes beyond the ${\cal DT}_{2v}$ cases we have examined.
We show in Fig.~\ref{fig:linear_RT} a case similar to that of  Fig.~\ref{fig:linear} but with $g(\vec{r})$ increasing clockwise from $-1$ at $\phi=0$ to 0 at $\phi=-\pi$ instead. It is described by a cyclic group of order 2 (i.e. $\{\mathbb{1},{\cal RT}\}$), and the system maintains a real spectrum until $\tau$ surpasses $\tauEP\simeq0.22$, because of the $\cal RT$ symmetry as discussed above. It is possible to apply the group theoretical analysis we have done above for ${\cal DT}_{2v}$ to generalized cyclic groups ${\cal CT}_{2v}$ ($v$ is an integer). The original cyclic group $C_{2v}$ we consider here consist of $2v$ rotations, which can be ordered by ${\cal R}_\frac{\pi}{v}$, ${\cal R}_\frac{2\pi}{v}$, \ldots, ${\cal R}_{2\pi}(=\mathbb{1})$, and ${\cal R}_\frac{\pi}{v}$ is the generator of the group. In the generalized cyclic group, half of these rotation operators, at the odd positions in the given order, are multiplied by $\cal T$, and ${\cal R}_\frac{\pi}{v}{\cal T}$ becomes the generator of the group. For example, ${\cal CT}_{4} = \{{\cal R}_\frac{\pi}{2}{\cal T},{\cal R},{\cal R}_\frac{3\pi}{2}{\cal T},\mathbb{1}\}$ and
${\cal CT}_{6} = \{{\cal R}_\frac{\pi}{3}{\cal T},{\cal R}_\frac{2\pi}{3},{\cal RT},{\cal R}_\frac{4\pi}{3},{\cal R}_\frac{5\pi}{3}{\cal T},\mathbb{1}\}$. ${\cal CT}_{2v}$ ($v$ is odd) contains $\cal RT$ and gives a protected symmetric phase at infinitesimal $\cal T$-breaking. For the rest of ${\cal CT}_{2v}$ (i.e. $v$ is even), they do not include $\cal RT$ and a partial transition takes place, similar to that shown in Fig.~\ref{fig:PTwheel}.

\begin{figure}[h]
\begin{center}
\includegraphics[width=0.8\linewidth]{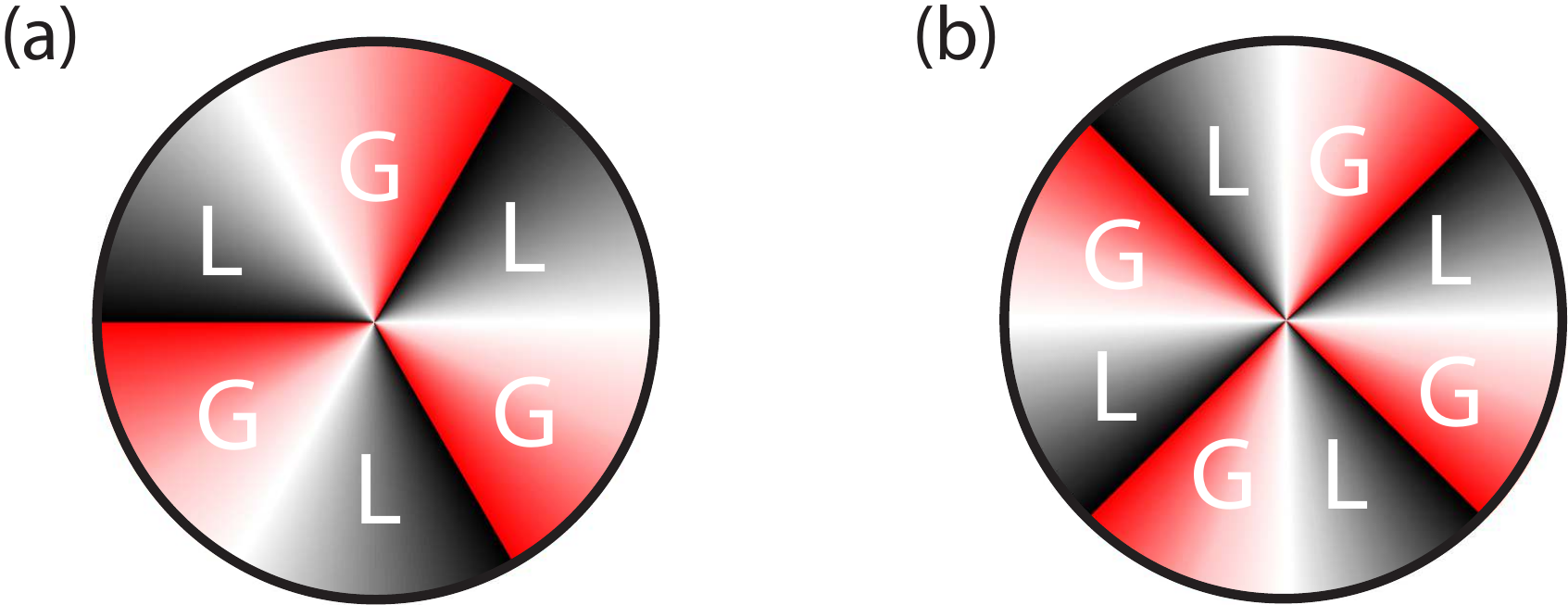}
\caption{(Color online) Examples of the ${\cal MT}_{2v}$ group with $v=3$ (a) and $v=4$ (b). Similar to the schematics in Figs.~\ref{fig:linear} and \ref{fig:linear_RT}, black and red represent loss (``$L$") and gain (``$G$") and the transparency of the colors show their strength.
}
\label{fig:MT}
\end{center}
\end{figure}

We note that for all generalized point groups with $\cal T$-breaking, there are always half of the elements that contain $\cal T$. This observation can be proved rigorously using the closure property of a group, which holds in higher dimensions as well. This finding serves as a guideline to find other generalized point groups with $\cal T$-breaking. For example, another way of generalizing the dihedral group ${\cal D}_{2v}$ with $\cal PT$ symmetry is multiplying all the $v$ reflection operators in it by $\cal T$. Since ${\cal P}_\frac{\pi}{2}$ and ${\cal RT}$ are not among the resulting elements, the spectrum always becomes (partially) complex at infinitesimal $\tau$. Two examples of this group, denoted by ${\cal MT}_{2v}$ with $v=3,4$, are given in Fig.~\ref{fig:MT}. They each have $2v$ angular blocks, instead of $v$ as in ${\cal DT}_{2v}$. We also note that we cannot simply multiply each element in a generalized point group by $\cal T$ to get another one. Although this procedure does not violate the guideline mentioned above, it removes $\mathbb{1}$ which is required by any group. We leave more complicated groups $\cal S$ derived from higher symmetry objects with imposed $\cal PT$ perturbations for future work.  We will however discuss the case of the sphere briefly below.

\section{3D sphere: odd degeneracy}

So far we have assumed pairwise degeneracy. What happens in the case of odd degeneracy? Our general arguments presented above suggest that either all degenerate eigenvalues will become complex for an infinitesimal value of $\tau$, or an even number will be protected by additional symmetries, leaving at least some fraction of the spectrum vulnerable to infinitesimal perturbations.
Here we analyze the scalar modes in a uniform sphere as an example. The hermitian eigenmodes at $\tau=0$ of each angular momentum $l$ and radial quantum number $\eta$ can be expressed as $j_\eta(nk_{l,\eta}r)Y_{l,m}(\theta,\phi)$, which have a $(2l+1)$-fold degeneracy. Here $m$ is the $z$-component of $l$, which we choose to be in the vertical direction, and $j_\eta,Y_{l,m}$ are the spherical Bessel function and the spherical Harmonics, respectively. We define the $\cal PT$-symmetry with respect to the $\phi=0$ plane, and for convenience we choose our basis as the parity eigenmodes about this plane, i.e. use $Y_{l,0}(\theta,\phi), Y_{l,m}^{(e)}\equiv[Y_{l,m}+(-1)^mY_{l,-m}]/\sqrt{2}\propto\cos(m\phi), Y_{l,m}^{(o)}\equiv[Y_{l,m}-(-1)^mY_{l,-m}]/\sqrt{2}i\propto\sin(m\phi)$ to denote their angular dependency. Since $Y_{l,0}(\theta,\phi)$ is uniform in $\phi$ and thus even about the $\cal PT$-plane, it couples to $Y_{l,m}^{(o)}$ but not $Y_{l,m}^{(e)}$. For the simplest non-trivial case with $l=1$, Eq.~(\ref{eq:modeCoupling}) can be approximated by the following $3\times3$ form:
\be
\begin{pmatrix}
1 & i\ttau G_{eo} & 0\\
i\ttau G_{eo} & 1 & i\ttau G_{o0}\\
0 & i b \ttau G_{o0} & 1
\end{pmatrix}
\begin{pmatrix}
a_e\\
a_o\\
a_0
\end{pmatrix}
=
\frac{k_{l,\eta}^2}{\tk^2}
\begin{pmatrix}
a_e\\
a_o\\
a_0
\end{pmatrix}.\label{eq:modeCoupling3x3}
\ee
Note that the degeneracy reduces the diagonal matrix $\bm{E}$ on right hand side of Eq.~(\ref{eq:modeCoupling}) to a scalar $k_{l,\eta}^2$. This results in a profound influence on the multimode coupling: one of the resulting eigenfrequeices stays unchanged and real, and its wave function is given by $G_{o0}Y_{l,1}^{(e)} - G_{eo}Y_{l,0}$ with a proper normalization. Note that this mixing is between the two hermitian modes that do not couple directly; they indirectly couple via $Y_{l,1}^{(o)}$, which however does not appear in the mixing. Thus this mode is the analogue of a ``dark state'' in atomic physics, and its form does not depend on the specific values of the nonvanishing coupling $G_{o0}, G_{eo}$. The other two $\tk$ are complex conjugates, whose imaginary parts again display a linear dependence on the $\cal T$-breaking:
\be
\im[\tk_{1,2}] \approx \pm\frac{k_{l,\eta}\sqrt{G_{eo}^2+G_{o0}^2}}{2}\,\ttau. \label{eq:imkR_3D}
\ee
All three hermitian modes are present in the two corresponding non-hermitian modes, which resemble the ``bright states'' but with complex eigenvalues.
We find similar results for $l>1$, and we therefore conclude that in this case of odd degeneracy the standard $\cal PT$-phase transition is also absent, but one
has a partial $\cal PT$-breaking transition, with a different fraction of the degenerate spectrum undergoing the transition for different values of $l$.

\section{Multimode coupling: restoring the $\cal PT$-symmetric phase at finite $\tau$}
\label{sec:multimode}
\begin{figure}[t]
\begin{center}
\includegraphics[width=\linewidth]{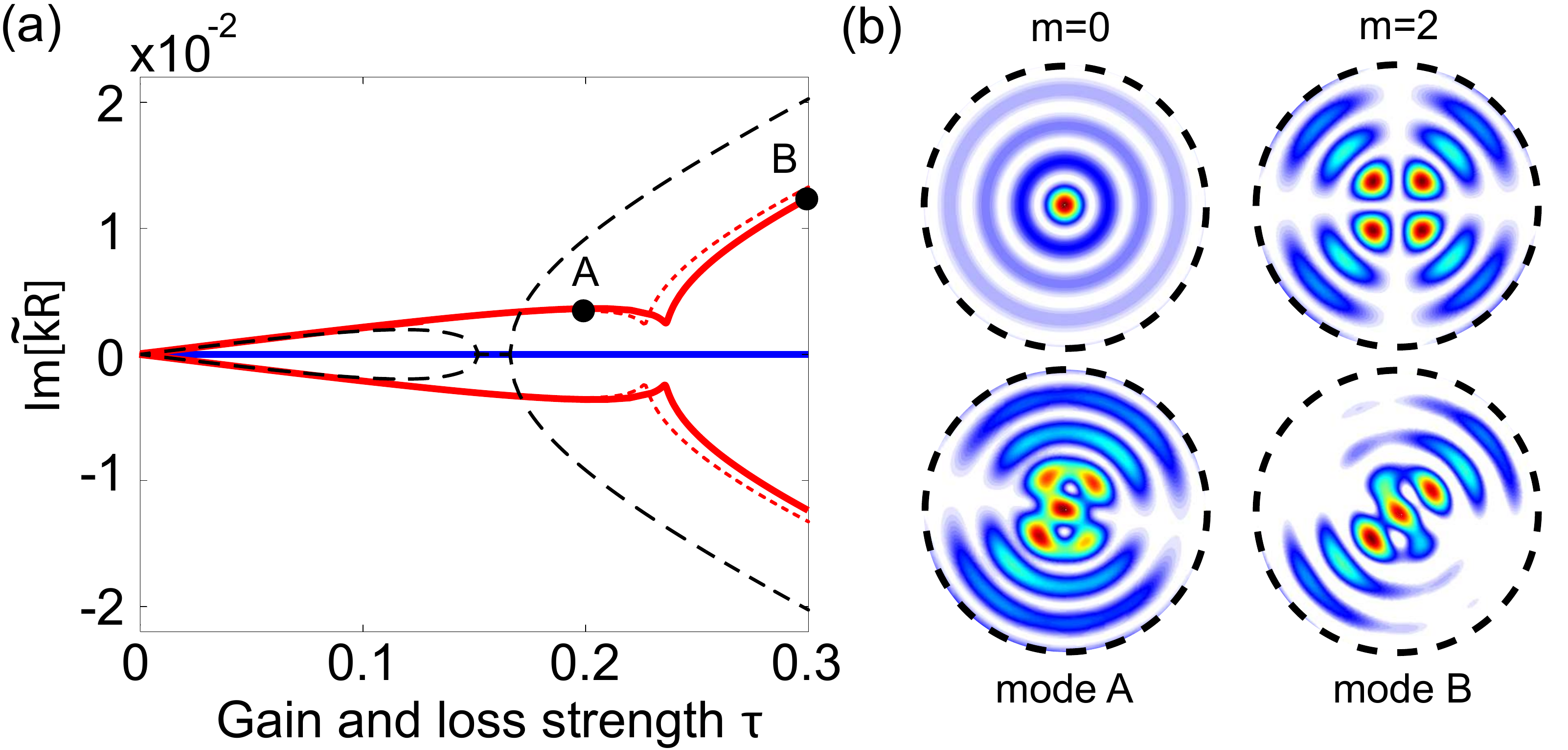}
\caption{(Color online) Restoration of the $\cal PT$-symmetric phase due to multimode coupling. (a) Evolution of the imaginary part of two $m=2$ and one $m=0$ eigenfrequencies as a function of the gain and loss strength $\tau$ (solid lines) in the $\cal{PT}$-symmetric disk studied in Fig.~\ref{fig:linear}. $k_0R=2.62$ and $k_2R=2.55$ at $\tau=0$.
Dotted lines show the three-mode approximation (\ref{eq:modeCoupling3x3}). Dashed lines show the case if the coupling between the $m=0$ mode and the odd-parity $m=2$ mode increased by $30\%$, which exhibits a restored $\cal PT$-symmetric phase in $\tau\in[0.152,0.166]$.
(b) Top: Intensity profiles of the $m=0$ mode and the odd-parity $m=2$ mode
at $\tau=0$. Bottom: The $\cal PT$-symmetry broken modes A and B at $\tau=0.2,0.3$ [marked by black dots in (a)]. Only the central regions of $r<R/2$ are shown. }
\label{fig:linear_multipole}
\end{center}
\end{figure}

Multimode coupling can be important even in the case of even degeneracy. We note that some eigenfrequencies in our 2D circular systems exhibit a more complicated dependence on $\tau$ in the $\cal PT$-broken phase [see the solid lines in Fig.~\ref{fig:linear_multipole}(a)].
Such nonmonotonic behaviors are signatures of multimode coupling, and in the case shown in Fig.~\ref{fig:linear_multipole}(a) they are primarily due to the couplings of two degenerate $m=2$ modes and the $m=0$ mode closest to them in frequency, which we denote $\wf^{(e)}_2,\wf^{(o)}_2,\wf_0$ and the corresponding eigenfrequencies $k_2,k_0$. $\wf_0$ contributes a significant fraction to the broken symmetry modes, as can be seen from their intensity profiles near the origin in Fig.~\ref{fig:linear_multipole}(b).
The $\phi$-dependence of these three modes is similar to the three $l=1$ modes analyzed in the sphere, which leads to a vanished $G_{e0}$ (not $G_{eo}$) and a similar $3\times3$ coupling matrix to that in Eq.~(\ref{eq:modeCoupling3x3}). As a result, two eigenmodes are in the broken-symmetry phase with infinitesimal $\cal T$-breaking and the other one stays in the symmetric phase. The existence of the latter is determined solely by the fact that these non-hermitian eigenfrequencies are given by the roots of a cubic equation with {\it real} coefficients. The latter is well known but has never found its way into non-hermitian systems as far as we know, since it is very rare to find real eigenvalues systematically in non-hermitian systems without the $\cal PT$-symmetry. These $\cal PT$-symmetric modes are no longer dark modes though; the bottom right diagonal element in the coupling matrix now becomes $k_2^2/k^2_0$ instead of unity, and as a result, $\wf^{(e)}_2,\wf^{(o)}_2,\wf_0$ are present in all three resulting non-hermitian modes. We further note that the ``gulfs'' near $\tau=0.23$ can develop into a restored $\cal PT$-symmetric phase, if the $\tau$-dependent discriminant of the aforementioned cubic equation becomes greater than zero. This condition can be satisfied, for example, if $G_{o0}$ increases by more than $5\%$; at $|G_{o0}/G_{eo}|\simeq3.57$, this restored $\cal PT$-symmetric phase is located at $\tau\in[0.152,0.166]$ [see the dashed line in Fig.~\ref{fig:linear_multipole}(a)].

\section{Discussion and conclusions}

Degeneracies in higher-dimensional systems are likely to be imperfect, especially for fabricated devices in photonics. This will lead to a finite but small $\cal PT$-transition threshold $\tauEP$. Since it is also very difficult to probe the system near $\tau=0$, the results we have presented hold qualitatively as long as $\tauEP$ is lower than the minimum $\tau$ reachable. In fact, in the presence of a global or a local defect satisfying the $\cal PT$-symmetry, one can tune the system across the symmetry transition threshold by controlling the amplitude of the defect. This scheme may benefit certain applications of $\cal PT$-symmetry based devices, such as sensing and switching.
This approach to a quasi-degenerate spectrum at $\tau=0$ can be regarded as coupling localized modes in momentum space; another approach is coupling two localized modes in the real space, and one example is the one-dimensional $\cal PT$-symmetric random chain studied in Ref.~\cite{Bendix}.
We also note that in Ref. \cite{Feng} the authors found a thresholdless condition for asymmetric modal conversion in quasi-1D waveguides with longitudinal index modulation $\Delta n(z)$ and attributed it to the spontaneous breaking of ${\cal PT}$-symmetry. However this situation was first studied by Greenberg and Orenstein \cite{Greenberg1,Greenberg2}, whose analysis makes it clear that this phenomenon is not related to ${\cal PT}$ symmetry breaking in our sense.

In summary, we have shown that the standard $\cal PT$-transition is absent for systems with degeneracies in the absence of $\cal T$-breaking as long as the perturbations are generic, i.e. without preserving additional discrete symmetries. As a result, a real spectrum cannot be maintained even at infinitesimal strength of gain and loss. If the $\cal T$-breaking also preserves additional discrete symmetries, the standard $\cal PT$-transition can be restored for certain subsets of the degenerate spectrum or even its entirety. We have applied a coupled mode theory to analyze these behaviors and given a group theoretical description of the ``protected" transitions, which also shows that qualitatively different behaviors can take place in $\cal PT$ and $\cal RT$ symmetric systems.

\section*{Acknowledgement}

We thank Konstantinos Makris, Ramy El-Ganainy, Stefan Rotter, and Jan Wiersig for helpful discussions. This project was partially supported by PSC-CUNY 45 Research Award and NSF under Grant No. ECCS 1068642.

\bibliographystyle{longbibliography}

\begin{thebibliography}{100}
\bibitem{smatrix_deg} One difference for the scattering matrix is the existence of a trivial degeneracy that occurs in waveguides if there is no scattering before the $\cal T$-breaking perturbation is turned on.
\end{thebibliography}

\end{document}